\documentclass[runningheads]{llncs}
\usepackage{graphicx}
\usepackage{comment}
\usepackage{pifont}
\usepackage{subcaption}
\usepackage{caption}
\newcommand{\xmark}{\ding{55}}%
\usepackage{multirow}
\usepackage{gensymb}
\newcommand{\blue}[1]{\textcolor{blue}{#1}}
\newcommand{\red}[1]{\textcolor{red}{#1}}

\newcommand\Tstrut{\rule{0pt}{2.5ex}}
\newcommand\Bstrut{\rule[-1ex]{0pt}{0pt}}

\newcommand{\widthscalefive}{0.145}
\usepackage{adjustbox}
\def\HS{\hspace{\fontdimen2\font}}
\def\HSE{\HS\HS\HS\HS\HS\HS\HS\HS}

\usepackage{array,xcolor,colortbl}
\newcolumntype{P}[1]{>{\centering\arraybackslash}p{#1}}

\usepackage{amsmath,amssymb} 
\usepackage{color}

\begin{document}
\pagestyle{headings}
\mainmatter
\def\ECCVSubNumber{111}  

\title{Multi-Attention Based Ultra Lightweight Image Super-Resolution} 

\titlerunning{Multi-Attention Based Ultra Lightweight Image Super-Resolution}
%

\newcommand*\samethanks[1][\value{footnote}]{\footnotemark[#1]}
\author{Abdul Muqeet\inst{1} \and
Jiwon Hwang\inst{1}\and
Subin Yang\inst{1}\and
JungHeum Kang\inst{1} \and
Yongwoo Kim\inst{2}\thanks{co-corresponding authors} \and
Sung-Ho Bae\inst{1}\samethanks}%
\authorrunning{A. Muqeet et al.}
%



\institute{Dept. of Computer Science and Engineering, Kyung Hee University, South Korea 
\email{\{amuqeet,jiwon.hwang,ysb8049,chhkang123,shbae\}@khu.ac.kr} \and
Dept. of System Semiconductor Engineering, Sangmyung University, South Korea \\
\email{yongwoo.kim@smu.ac.kr}}
\maketitle

\begin{abstract}

Lightweight image super-resolution (SR) networks have the utmost significance for  real-world applications. There are several deep learning based SR methods with remarkable performance, but their memory and computational cost are hindrances in practical usage. To tackle this problem, we propose a Multi-Attentive Feature Fusion Super-Resolution Network (MAFFSRN). MAFFSRN consists of proposed feature fusion groups (FFGs) that serve as a feature extraction block. 
Each FFG contains a stack of proposed multi-attention blocks (MAB) that are combined in a novel feature fusion structure. Further, the MAB with a cost-efficient attention mechanism (CEA) helps us to refine and extract the features using multiple attention mechanisms. The comprehensive experiments show the superiority of our model over the existing state-of-the-art. We participated in AIM 2020 efficient SR challenge with our MAFFSRN model and won 1st, 3rd, and 4th places in memory usage, floating-point operations (FLOPs) and number of parameters, respectively. 


\end{abstract}

\begin{figure}[t]
\centering
  \includegraphics[width=0.999\linewidth]{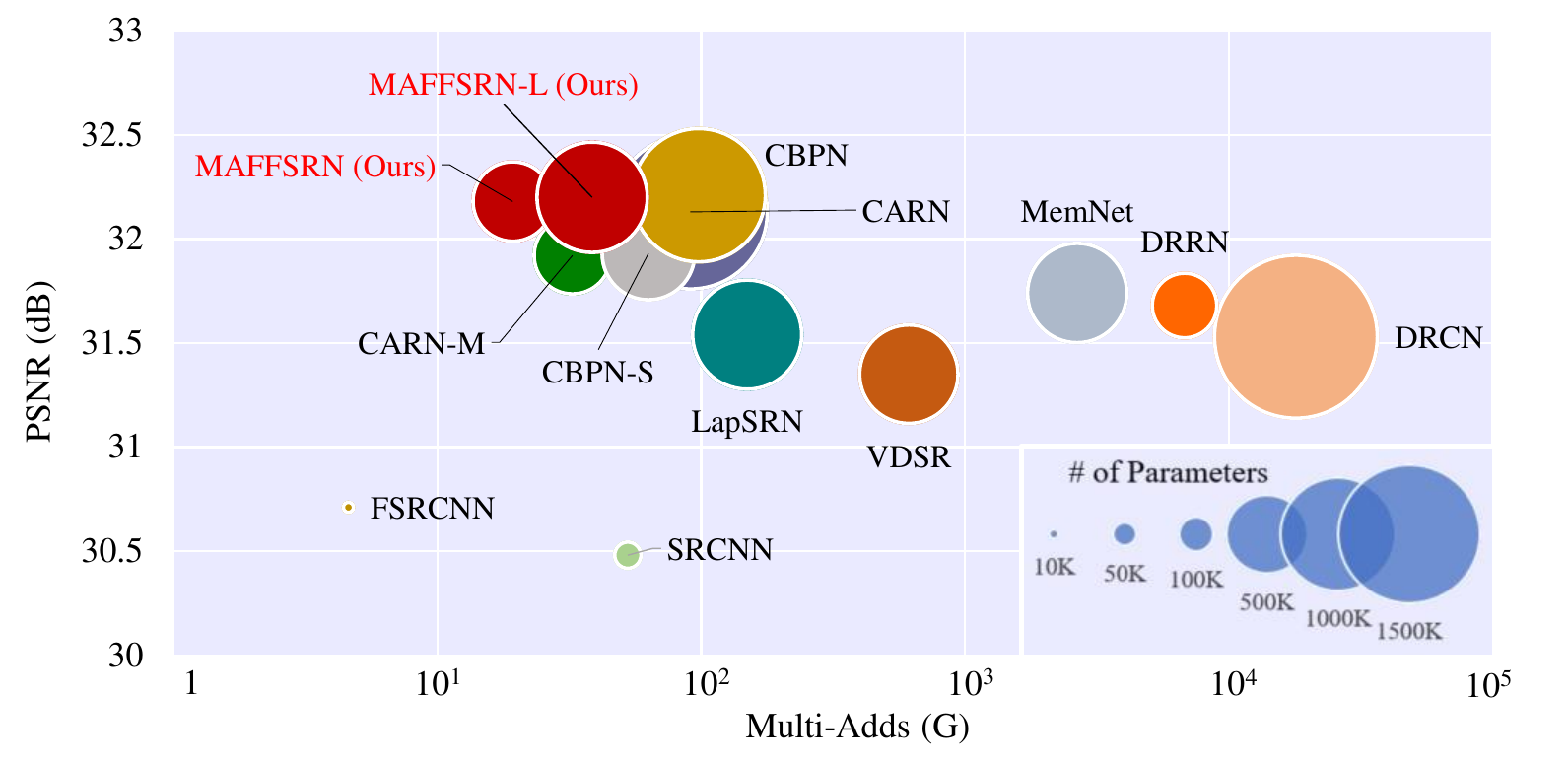}
  \caption{Performance comparison of existing lightweight methods on Set5 \cite{bevilacqua2012low} (4$\times$). Multi-adds are calculated on 720p HR image. The results show the superiority of our models among existing methods }
  \label{fig:efficient_fig}
\end{figure}

\section{Introduction}

This paper focuses on the single image super-resolution (SISR) problem. In SISR we aim to reconstruct a high-resolution (HR) image from a low-resolution (LR) image. We refer super-resolution (SR) as interchangeably with SISR in remaining of the paper. According to \cite{zhang2020deep}, SISR problem can be mathematically written as,

\begin{equation}\label{eq:sisr_degradation}
  I_{LR}\!= \!(I_{HR}\otimes k)\!\downarrow_{\bf{s}}\! + n,
\end{equation}

\noindent where $I_{LR}$ and $I_{HR}$ refer to given input LR and desire HR images. `\textit{k}' in Eq. \ref{eq:sisr_degradation}) denotes as a blur kernel, ${ \downarrow_s}$ represents a down-scaling operator, and `\textit{n}' is a Gaussian noise. By following previous works, we assume that images are down-sampled with bicubic interpolation \cite{ahn2018fast}\cite{zhu2019efficient}.

There are multiple mapping solutions possible from single LR to HR images that make this problem ill-posed. Albeit of its ill-posedness, the deep learning methods like \cite{dong2015image,dong2016accelerating,Kim_2016_DRCN} have shown notable success in this domain. For instance, SRCNN \cite{dong2015image} with only three layers outperformed the previous non-deep learning methods. Subsequently, deeper and complex architectures have been proposed to improve the performance of SR methods \cite{ahn2018fast,lim2017enhanced,haris2018deep,zhang2018image,zhang2018residual}. In spite of their outstanding performance, such methods are impracticable for real-world applications because of their large memory size, number of operations, and parameters.

Numerous lightweight models have been proposed to resolve these issues. CARN \cite{ahn2018fast} introduces a lightweight and efficient cascaded residual network with several residual connections. FALSR \cite{chu2019fast} employs a network architecture search (NAS) technique rather than manually searching it in SISR domain. CBPN \cite{zhu2019efficient} proposed an efficient version of the DBPN network \cite{haris2018deep} that emphasizes the importance of high-resolution features of LR images. These models were designed to reduce the computational cost,  though all of these come with their smaller version of models, such as, CARN-M\cite{ahn2018fast}, CBPN-S\cite{zhu2019efficient}, FALSR-B, and FALSR-C\cite{chu2019fast}. Hence, it shows that their original models are inadequate for real-world application.

The need of such practical models motivated us to propose a lightweight model called MAFFSRN. Its computation cost is similar to CARN-M, CBPN-S, FALSR-B, FALSR-C  \cite{ahn2018fast,zhu2019efficient,chu2019fast}, but matches the performance of their corresponding original models. With the comprehensive experiments, we show that our models achieve the best performance on all of the benchmark datasets.  Further, we introduce a large model (MAFFSRN-L) to compare performance with heavy state-of-the-art methods. Note that its size still remains smaller than existing efficient models. 
We show our benchmark results in Figure \ref{fig:efficient_fig}.

Our model is specifically aimed to minimize the computation cost such as floating point operations (FLOPs) and memory consumption but maximize the network performance. To increase the network performance, we utilize the feature fusion group (FFG) that consists of several multi-attention blocks (MAB). In SR deep network architectures the vital information is vanished during the flow of network \cite{zhang2018image}.
Our method tackles this problem with FFG and MAB and results suggest that they enable us to increase the depth of network with minimal computational cost, consequently increasing the  network performance.  
The next challenge is to minimize the computational cost and memory usage. For this purpose, we propose changes for the enhanced spatial attention (ESA) block \cite{liu2020residual}. First, we introduce cost-efficient (CEA) block to directly apply attention mechanism on the input features. Second, we replaced the Conv groups of ESA \cite{liu2020residual} with dilated convolutions to get benefit from the large spatial size. For the feature fusion structure, we found during the experiments that the performance of hierarchical feature fusion (HFF) \cite{li2018multi} remains lower than the  binarized feature fusion (BFF) \cite{muqeet2019hybrid}. We discuss the details of these experiments in ablations studies. We evaluate our method on benchmark datasets and compare the performance against existing methods. 

Our overall contributions are summarized as follows: 1)  we introduce a lightweight model consisting of modified BFF, MAB, and CEA modules, that outperforms existing methods. We participated in AIM 2020 SR challenge \cite{zhang2020aim} where our model was ranked $1^{st}$ in memory consumption, $3^{rd}$ in FLOPS, and $4^{th}$ in number of parameters. 2) We provide comprehensive qualitative and quantitative comparison results on the benchmark datasets with multiple scaling factors ($\times$2, $\times$3, and $\times$4).   




\begin{figure}[t]
\centering
  \includegraphics[width=0.999\linewidth]{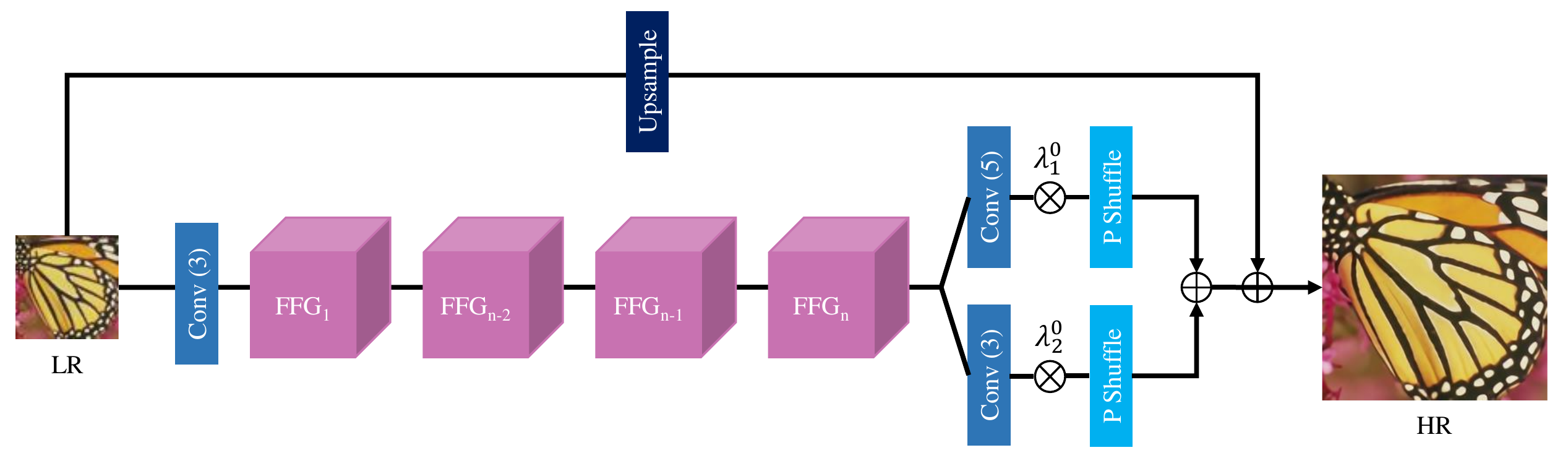}
  \caption{Our main network architecture (MAFFSRN). It consists of stack of FFG where each FFG has multiple MAB combined with modified BFF. Conv ($3$) and Conv ($5$) refer to  $3\times3$ and $5\times5$ convolutions, respectively,P Shuffle means pixel shuffle \cite{shi2016real}. Each lambda ($\lambda$) is trainable scalar  parameter. Lastly, we add  up-sampled LR image to the reconstructed output }
  \label{fig:main_fig}
\end{figure}
\section{Related work} 

The remarkable success of deep neural networks in other computer vision tasks \cite{hu2018squeeze},\cite{he2016deep},\cite{chollet2017xception} encouraged SR community to apply deep learning techniques in SR domain. SRCNN \cite{dong2015image} apply a shallow neural network and surpasses the performance of traditional and conventional non-deep learning based methods. As \cite{urban2016deep} shows that deep networks have shown better results than shallow networks, several methods followed this trend and proposed deeper networks. VDSR \cite{kim2016accurate} proposed a 20 layers network consisting of global skip-connection that element-wise add the up-sampled LR image to the output reconstructed image. EDSR \cite{lim2017enhanced} improved the SRResNet \cite{ledig2017photo}, that was based on ResNet architecture\cite{he2016deep}, by removing the trivial layers or those layers which degrade the performance, such as Batch Normalization\cite{ioffe2015batch}. RDN introduced dense connections similar to DenseNet \cite{huang2017densely} and improved the performance with fewer parameters than EDSR\cite{lim2017enhanced}. Certainly, they have improved the image fidelity, such as PSNR or SSIM significantly, however, the constrained real-world environments having low-power computing devices require to focus on other metrics, such as number of parameters, memory consumption, FLOPs, latency time, etc.   

Therefore, there is growing interest to build lightweight models that need to be accurate as well. One strategy is to adopt model compression techniques to compress the models \cite{han2015deep,hinton2015distilling}. In this paper, our focus is to develop a new network architecture to remedy this problem. Hence, we only discuss the previous works that address such issues in SR domain.


The progress of such lightweight architectures started from FSRCNN \cite{dong2016accelerating}. It improves the performance of SRCNN \cite{dong2015image} by directly applying SR network to LR images rather than up-sampled input. It also decreases the inference time by removing the high-cost up-sampling layers. DRRN \cite{Tai-DRRN-2017} utilized recursive layers to reduce the number of parameters while keeping the depth of the network. CARN \cite{ahn2018fast} applied several residual connections and recursive layers to reduce computational cost. FALSR \cite{chu2019fast} introduced automated neural architecture search (NAS) strategies in SR domain to propose an SR model for constrained environment. 
CBPN \cite{zhu2019efficient} proposed an efficient version of DBPN network \cite{haris2018deep} by replacing the expensive up- and down- projection modules with pixel shuffle layers. We observed that all these methods focused on trade-off between performance and computation cost that led them to propose another smaller version of their model such as CBPN-S \cite{zhu2019efficient}, CARN-M \cite{ahn2018fast}, FALSR-B \cite{chu2019fast}. However, our proposed method achieves better or comparable performance as their original models whereas the computational cost remains the same or lower compared to their lightweight versions.

\section{Proposed Method}

In this section, we describe the details of our proposed architecture. As shown in Figure \ref{fig:main_fig}, our network architecture consists of \textit{n} FFGs that are stacked in a sequential way. The details of FFG are given section \ref{section:FFG}. We use one convolutional (Conv) layer before FFGs to extract the shallow features from input LR image. Lastly, we apply couple of Conv layers with different filter sizes to extract multi-scale features that are followed by pixel-shuffle layers \cite{shi2016real}. Further, motivated from \cite{wang2019lightweight}, we add weights (denoted by $\lambda^0_{1}$ and $\lambda^0_{2}$) to both Conv layers to give weightage to the features and carry the weighted features to later layers. Later, the resultant information is element-wise added. Similar to \cite{kim2016accurate}, we element-wise add up-sample LR input into the output layer. Note, our overall architecture is primarily based on RDN architecture \cite{zhang2018residual} that consists of local and global blocks. 

For the given I$_{LR}$ image, the shallow feature extraction step is given as

\begin{equation}
x_{sfe}=f_{sfe}\left(I_{LR}\right),
\end{equation}

\noindent where $f_{sfe}$ and $x_{sfe}$ represent the $3\times3$ convolution and the resultant output, respectively. Next, for non-linear mapping or deep feature extraction step, we apply the stack of FFG as follows 

\begin{equation}
x_{dfe}=f_{FFG}^{n}\left(f_{FFG}^{n-1}\left(\ldots f_{FFG}^{0}\left(x_{sfe}\right) \right)\right),
\end{equation}

\noindent where $f_{FFG}^{n}$ and $x_{dfe}$ denote the \textit{n$_{th}$} FFG  and output of deep feature extraction step, respectively. Lastly, the reconstruction stage is given as

\begin{equation}\label{eq:sr}
I_{SR}=  f_{ps}\left( \lambda^0_1 f_{5}\left(x_{dfe}\right)\right) + f_{ps}\left( \lambda^0_2 f_{3}\left(x_{dfe}\right)\right) + f_{up}\left(I_{LR}\right),
\end{equation}

\noindent where the details of notations in Eq. \ref{eq:sr} are as follows: $I_{SR}$ represents the desired SR image, $f_{3}$ and $f_{5}$ denote $3\times3$ and $5\times5$ convolutions, respectively, $f_{ps}$ shows pixel-shuffle layer \cite{shi2016real}, $f_{up}$ represent an up-sampling layer and  $\lambda^0_1$ and $\lambda^0_2$ denote trainable scalar parameters.

\begin{figure}[!t]
  \includegraphics[width=\linewidth]{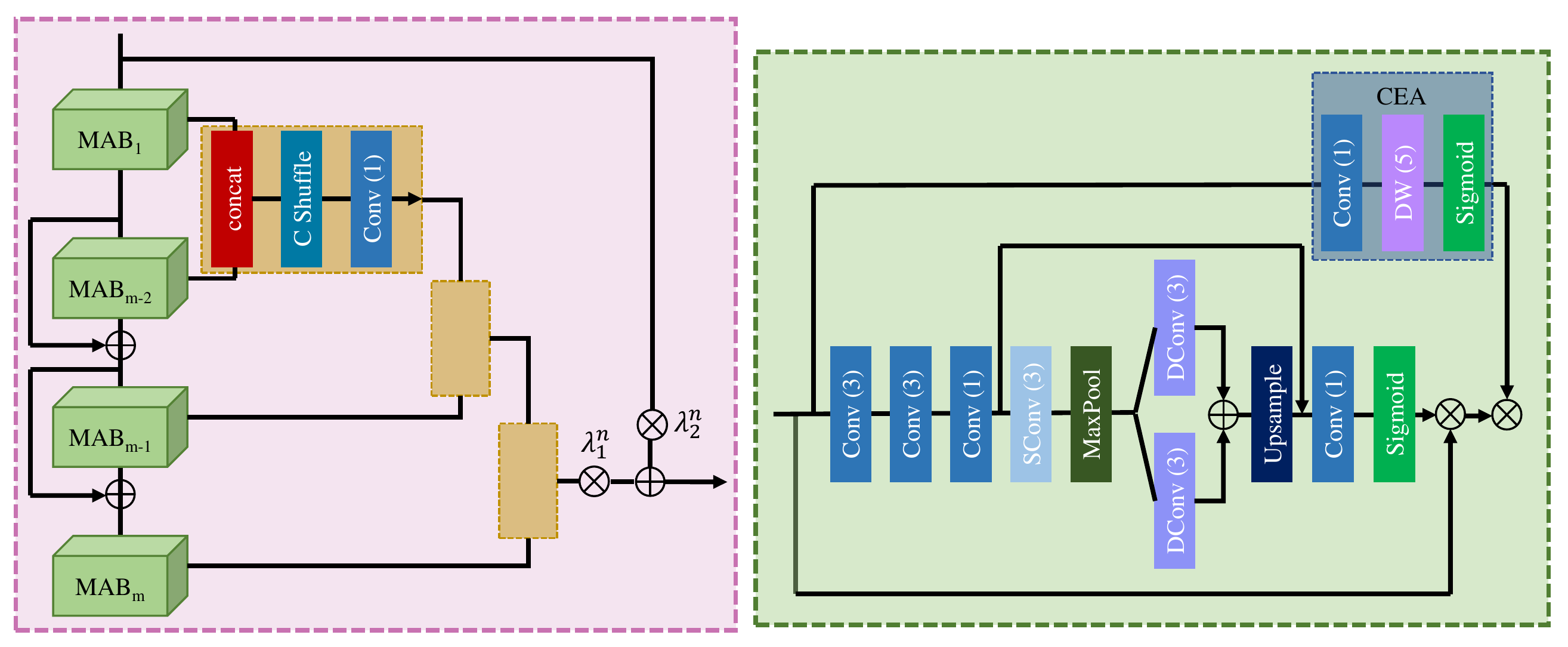}
  \caption{Left figure shows structure of FFG and right figure shows MAB. `C Shuffle' refers to channel shuffle, `SConv' means strided convolution, `DConv' means dilated convolution and `DW' means depth-wise convolution. Parenthesis such as (1), (3) and (5) represents 1$\times$1, 3$\times3$, and 5$\times5$ filters, respectively}
  \label{fig:local_group_blocks}
\end{figure}

\begin{figure}[!t]
  \includegraphics[width=\linewidth]{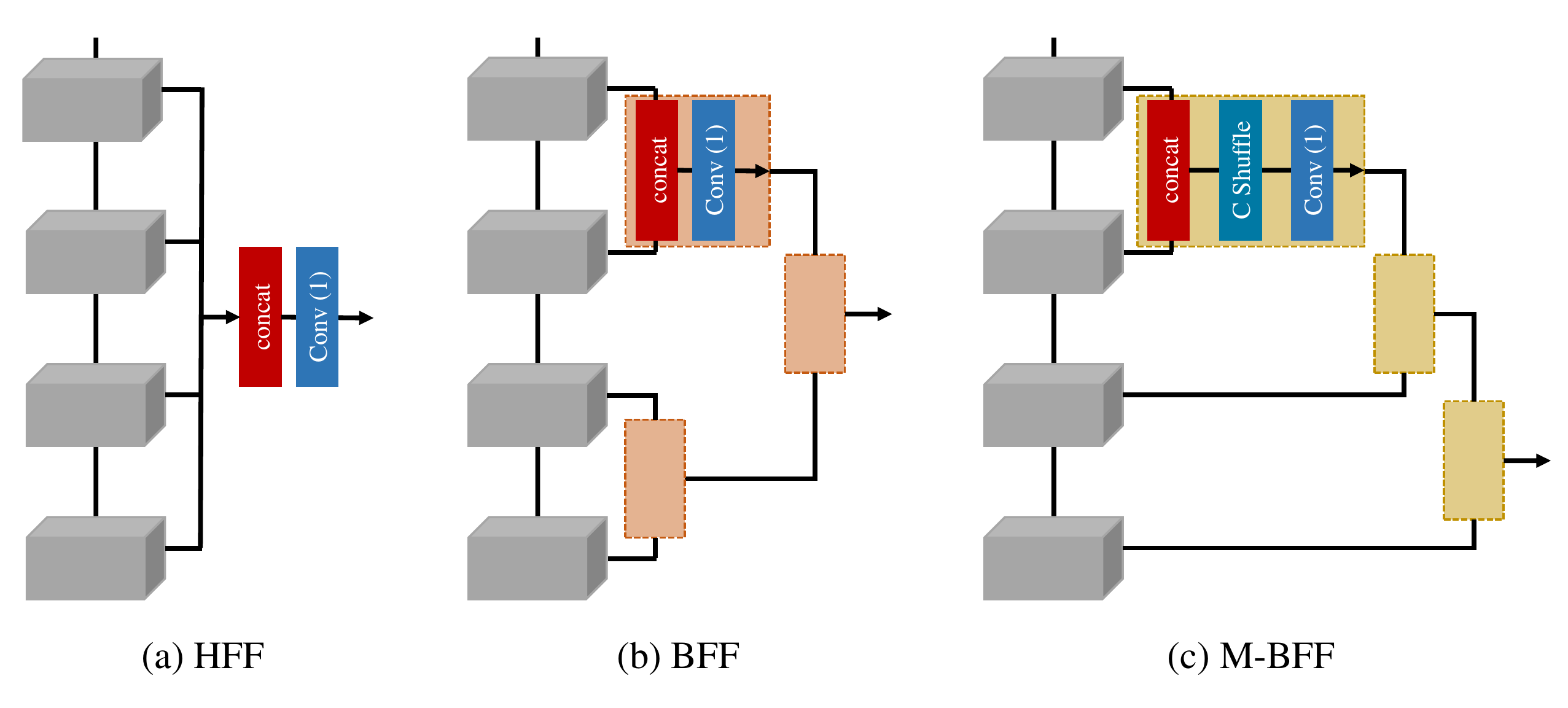}
  \caption{Comparison of different feature fusion structures (1) HFF \cite{Li_2018_ECCV}, (2) BFF \cite{muqeet2019hybrid} , (3) M-BFF (Ours). In figure  `concat' refers to channel-wise concatenation, `Conv(1) refers to $1\times1$ convolution, and `C Shuffle' shows channel shuffle }
  \label{fig:FFS}
\end{figure}

\subsection{Feature Fusion Group (FFG)} \label{section:FFG}
Our proposed FFG has \textit{m} multi-attention blocks (MAB). The details of MAB are discussed in the next section. The proposed MABs are combined through a modified form of binarized feature fusion (BFF) structure \cite{muqeet2019hybrid}. HFF \cite{Li_2018_ECCV} is another fusion structure that is commonly used though during experiments we found that BFF \cite{muqeet2019hybrid} performs better than HFF \cite{Li_2018_ECCV} (details are discussed in section \ref{section:ablation}). We refer to modified BFF as M-BFF. The comparisons of structures are shown in Figure \ref{fig:FFS}. In BFF, all the adjacent blocks are separately concatenated like a binary tree structure. In contrast, our M-BFF concatenates the resultant feature block with the next MAB as shown on the left side of Figure \ref{fig:local_group_blocks}. By taking the inspiration from ShuffleNet \cite{zhang2018shufflenet} that applies a channel shuffle method to mix the information among groups, we introduce the channel shuffle to mix the information between concatenated channels followed by a channel reduction layer that reduces the channels to make it equal to the number of input channels. In the end, we element-wise add the input features to the output   features. Additionally, the residual connections may contain redundant information, thus to filter desired and useful information  we multiply the results with trainable parameters $\lambda^n_1$ and $\lambda^n_2$ where \textit{n} refers to \textit{n$^{th}$} FFG.



\subsection{Multi-attention Block (MAB)} 
In this section, we define the details of our proposed residual block, called MAB. \cite{zhang2018image} has emphasized the importance of channel attention (CA) mechanism. Consequently, many SR methods have focused on attention mechanisms, mainly CA and spatial attention (SA) \cite{woo2018cbam,kim2018ram,muqeet2019hybrid}. Recently,  \cite{liu2020residual} proposed a combined solution for CA and SA called enhanced spatial attention (ESA). The ESA  block reduces the number of channels with $1\times1$ convolutions and the number of spatial size with strided convolutions. Later, these spatial and channel sizes are increased to match the input size. Lastly,  sigmoid operation is applied to get a similar effect as channel attention mechanism \cite{hu2018squeeze}.

We modify the ESA block to make it more efficient by introducing dilated convolutions with different filter sizes. Further, we element-wise add all output features of dilated convolutions together to minimize the gridding effects \cite{yu2017dilated}. The dilated convolutions not only reduce our memory computations but also increase spatial filter sizes, enabling us to improve the performance. Further, we introduce another cost-efficient attention mechanism (CEA) \cite{cai2020learning} to refine our input features. The CEA consists of point-wise and depth-wise convolutions.  It is incorporated into the MAB block to improve the performance of our network with the negligible additional computational cost. The structure of MAB is presented on
the right side of Figure \ref{fig:local_group_blocks}.



\section{Experimental Setup}

\subsection*{ Implementation Details} 

As we focus on developing a lightweight model, we aim to maximize  the performance of existing networks as well as minimize their computational cost. 
We denote our original model as MAFFSRN.  Further, we also introduce our larger model MAFFSRN-L to show that we can enhance the performance of our model depending on the available computing resources. 

Our lightweight model MAFFSRN consists of 4 FFGs and 4 MABs whereas, for MAFFSRN-L model, we keep the same number of MABs and increase the FFGs to 8. 
We reduce the number of channels by a factor of 4 in MAB and set stride$=3$ to reduce the spatial size. The dilatation factors are set to $D=1$ and $D=2$. The values of scale $\lambda$ are initialized with 0.5. We set the number of filters to 32 for every Conv layer except the last layer. For the last layer, we use 3 filters to reconstruct 3-color images. It can be modified to 1 filter for grayscale images. 

\subsection*{Training Settings} 

We used AdamP optmizer \cite{heo2020slowing} to train our models with initial learning rate $2 \times 10^{-4}$. For the data augmentations, we apply standard techniques i.e., images are flipped horizontally or vertically and randomly rotated by 90\degree, 180\degree and 270\degree. The models are trained for 1000 epochs and learning rate is decreased to half after every 200 epochs. We set batch-size to 16 and input patch size to 48$\times$48.  We implement our network on PyTorch and train it on NVIDIA RX 2080TI GPU and select the best performance model.

\subsection*{Datasets}

We use the high-quality DIV2K \cite{agustsson2017ntire} dataset for training our models. It consists of LR and HR pairs of 800 training images. The LR images are obtained through bicubic down-sampling. For the evaluation of our models, we use the standard and publicly available benchmark datasets, Set5 \cite{bevilacqua2012low}, Set14 \cite{zeyde2010single}, B100 \cite{martin2001database}, and Urban100 \cite{huang2015single} datasets. Set5 \cite{bevilacqua2012low}, Set14 \cite{zeyde2010single}, B100 \cite{martin2001database} contain animals, people, and natural scenes, while Urban100 \cite{huang2015single} consists of urban scenes only.

\subsection*{Evaluation metrics}

We measure the performance of reconstructed SR images with PSNR and SSIM \cite{wang2004image} by following \cite{kim2016accurate}, using luminance or Y-channel of transformed YCbCr color space. We also calculate the number of parameters and multi-adds to compare the computational complexity of proposed models with existing methods.



\begin{table}[h]
\centering
\caption{Effects of BFF, M-BFF, and CEA modules. Experiments are performed on Set5 (2$\times$)}
\begin{tabular}{|ccccc|}
\hline
BFF& M-BFF & CEA & PSNR/SSIM &  Parameters\\ \hline
\xmark &  \xmark & \xmark    & 37.87/0.9601& 364K    \\ \hline
\checkmark &  \xmark & \xmark    & 37.91/0.9602 & 372K    \\ \hline
\xmark &\checkmark &  \xmark  &  37.94/0.9602&372K    \\ \hline
\xmark &  \xmark &\checkmark   & 37.93/0.9603& 394K      \\ \hline
\xmark &  \checkmark &\checkmark   & \textbf{37.97/0.9603}  &  402K     \\ \hline
\end{tabular}  \label{tab:ablation}

\end{table}

\subsection{Ablation Studies}  \label{section:ablation}
We conduct a series of ablation studies to demonstrate the importance of each proposed module used in our model. For all these experiments, we fully train our MAFFSRN model for 1000 epochs.  In the first experiment, we train multiple models with similar settings to show the overall contribution of M-BFF and CEA. Each time we remove one component and test the network performance without that specific module. The results are shown in Table \ref{tab:ablation}. It is noted that the model without M-BFF refers to model with HFF \cite{li2018multi} structure that is a common choice for SR methods. Row 2 of Table \ref{tab:ablation} suggests that M-BFF improves 0.07 dB PSNR with only 8K additional parameters. Similarly, CEA adds 0.05 dB with 32K parameters. Lastly, when we combine CEA and M-BFF, our model obtains 0.1 dB PSNR with less than 40K additional parameters. Note, for the fair comparisons, we add channel shuffle in all three methods.

\begin{table}[h]
\centering
\caption{Importance of channel shuffle in MAFFSRN. Experiments are performed on Set5 (2$\times$) }
\begin{tabular}{|c|c|}
\hline
Method                  & PSNR/SSIM \\ \hline

Without Channel Shuffle &   37.93/0.9603      \\ \hline
With Channel Shuffle (Ours)&  \textbf{37.97/0.9603}       \\ \hline

\end{tabular} \label{tab:ablation_shufflelayers}
\end{table}

\begin{table}[t]
\centering
\caption{Evaluation of performance of different Conv layers in MAB. Here `d' represents dilation factors. Experiments are performed on Set5 (2$\times$) with MAFFSRN }

\begin{tabular}{|c|c|c|c|}

\hline
Number   of Conv Layers & Details      & PSNR/SSIM & Params\\ \hline
2   (ours)                & 3x3, 3x3 (d=2)     &    \textbf{37.97}  /0.9603       &  402394   \\ \hline
2                       & 3x3, 3x3     &    37.95/0.9603  &  402394   \\ \hline
3                       & 3x3, 3x3,3x3 &    37.96/\textbf{0.9604}  &   411738   \\ \hline
2                       & 3x3, 5x5     &    37.89/0.9602  &   418778  \\ \hline
2                       & 5x5, 3x3     &    37.92/0.9602  &   418778  \\ \hline
\end{tabular} \label{tab:ablation_conv_layers}
\end{table}

To demonstrate the importance of channel shuffle in our proposed MAFFSRN, we remove channel shuffle from MAFFSRN and report the results in Table \ref{tab:ablation_shufflelayers}. The results clearly indicate that with channel shuffle, we can increase the performance up to 0.04 PSNR.

For further evaluation, we experimented with a different type of Conv layers to show the efficacy of dilated convolutions in MAB. The experimental results in Table \ref{tab:ablation_conv_layers} suggest that our dilated convolutions perform better results than three convolutions and, surprisingly, methods having $5\times5$ convolutions are the worst performances. The reason could be the structure of the Conv layers as element-wise addition of $3\times3$ and $5\times5$ Conv layers has no significant benefits over dilated convolutions that utilize both layers more effectively. 

We further experimented to compare the Adam \cite{kingma2014adam} and AdamP \cite{heo2020slowing} in Table \ref{tab:ablation_adamp} and found that AdamP \cite{heo2020slowing} consistently outperforms the Adam\cite{kingma2014adam} optimizer on all of the datasets with a large margin.

\begin{table}[!h]
\centering
 \caption{Performance comparison between Adam \cite{kingma2014adam} and AdamP\cite{heo2020slowing}}
\begin{tabular}{|c|cccc|}
\hline
Optimizer & Set5 \cite{bevilacqua2012low}      &  Set14\cite{zeyde2010single}         &  B100 \cite{martin2001database}     & Urban100 \cite{huang2015single}  \\ 
& PSNR/SSIM&PSNR/SSIM&PSNR/SSIM&PSNR/SSIM\\
\hline
Adam\cite{kingma2014adam} &37.87/0.9601& 33.42/0.9165  & 32.09/0.8987 & 31.75/0.9245          \\ 
AdamP\cite{heo2020slowing}  &  37.97/0.9603    & 33.49/0.9170  &  32.14/0.8994     &    31.96/0.9268 \\
\hline

\end{tabular}  \label{tab:ablation_adamp}

\end{table}




\begin{table}[!t]

\scriptsize
\setlength{\tabcolsep}{2pt}
\caption{Quantitative comparisons of existing methods on four datasets and three scales 2$\times$, 3$\times$, and 4$\times$. Red/blue/green text: best/second-best/third-best}
\scalebox{0.96}{%
\begin{tabular}{c l r r c c c c}
\hline\Tstrut
\multirow{2}{*}{Scale} & \multirow{2}{*}{Model} & \multirow{2}{*}{Params} & \multirow{2}{*}{Multi-Adds} & Set5 & Set14 & B100 & Urban100 \\
&  &  & & PSNR/SSIM & PSNR/SSIM & PSNR/SSIM & PSNR/SSIM
\Bstrut\\\hline\Tstrut
\multirow{12}{*}{2} 
& SRCNN\cite{dong2015image}    & 57K    & 52.7G    & 36.66/0.9542 & 32.42/0.9063 & 31.36/0.8879 & 29.50/0.8946 \\
& FSRCNN\cite{dong2016accelerating}  & 12K    & 6.0G     & 37.00/0.9558 & 32.63/0.9088 & 31.53/0.8920 & 29.88/0.9020 \\
& VDSR\cite{kim2016accurate}      & 665K   & 612.6G   & 37.53/0.9587 & 33.03/0.9124 & 31.90/0.8960 & 30.76/0.9140 \\
& DRCN\cite{Kim_2016_DRCN}      & 1,774K & 17,974.3G & 37.63/0.9588 & 33.04/0.9118 & 31.85/0.8942 & 30.75/0.9133 \\
& CNF\cite{cnf2017}        & 337K   & 311.0G   & 37.66/0.9590 & 33.38/0.9136 & 31.91/0.8962 & -\\
& LapSRN\cite{lai2017laplacian}  & 813K   & 29.9G    & 37.52/0.9590 & 33.08/0.9130 & 31.80/0.8950 & 30.41/0.9100 \\
& DRRN\cite{Tai-DRRN-2017}      & 297K   & 6,796.9G & 37.74/0.9591 & 33.23/0.9136 & 32.05/0.8973 & 31.23/0.9188 \\
& BTSRN\cite{fan2017balanced}      & 410K   & 207.7G   & 37.75/-\HSE  & 33.20/-\HSE  & 32.05/-\HSE  & 31.63/-\HSE \\
& MemNet\cite{Tai-DRRN-2017}      & 677K   & 2,662.4G   &37.78/0.9597 & 33.28/0.9142 & 32.08/0.8978 & 31.31/0.9195 \\
& SelNet \cite{choi2017deep}      & 974K   & 225.7G   & 37.89/\textcolor{green}{0.9598}  & \red{33.61}/0.9160  & 32.08/0.8984 & - \\
& FALSR-A\cite{chu2019fast}  & 1,021K & 234.7G   & 37.82/0.9595 & 33.55/0.9168 & 32.12/0.8987 & 31.93/0.9256 \\
& FALSR-B\cite{chu2019fast}    & 326K   & 74.7G    & 37.61/0.9585 & 33.29/0.9143 & 31.97/0.8967 & 31.28/0.9191 \\
& FALSR-C\cite{chu2019fast}   & 408K   & 93.7G    & 37.66/0.9586 & 33.26/0.9140 & 31.96/0.8965 & 31.24/0.9187 \\
& SRMDNF\cite{zhang2018learning}     &    &    & 37.79/0.9601 & 33.32/0.9159 & 32.05/0.8985 & 31.33/0.9204 \\
& CARN  \cite{ahn2018fast}            & 1,592K & 222.8G   & 37.76/0.9590 & 33.52/0.9166 & 32.09/0.8978& 31.92/0.9256 \\
& CARN-M  \cite{ahn2018fast}          & 412K   & 91.2G    & 37.53/0.9583 & 33.26/0.9141 & 31.92/0.8960 & 31.23/0.9193\\
& CBPN-S\cite{zhu2019efficient}     & 430K   & 101.5G   & 37.69/0.9583 & 33.36/0.9147 & 32.02/0.8972 & 31.55/0.9217 \\
& CBPN\cite{zhu2019efficient}       & 1,036K & 240.7G   &  \textcolor{green}{37.90}/0.9590 &  \blue{33.60}/\blue{0.9171} & \blue{32.17}/\textcolor{green}{0.8989} & \blue{32.14}/\blue{0.9279} \\
& MAFFSRN (ours)      & 402K   & 77.2G  & \blue{37.97}/\blue{0.9603} & 33.49/\textcolor{green}{0.9170} & \textcolor{green}{32.14}/\blue{0.8994} & \textcolor{green}{31.96}/\textcolor{green}{0.9268} \\
& MAFFSRN-L (ours)      & 790K   & 154.4G  & \red{38.07}/\red{0.9607} & \textcolor{green}{33.59}/\red{0.9177} & \red{32.23}/\red{0.9005} & \red{32.38}/\red{0.9308} \Bstrut \\\hline\Tstrut
\multirow{11}{*}{3} & SRCNN\cite{dong2015image}
                          & 57K    & 52.7G    & 32.75/0.9090 & 29.28/0.8209 & 28.41/0.7863 & 26.24/0.7989 \\
& FSRCNN\cite{dong2016accelerating}  & 12K    & 5.0G     & 33.16/0.9140 & 29.43/0.8242 & 28.53/0.7910 & 26.43/0.8080 \\
& VDSR\cite{kim2016accurate}      & 665K   & 612.6G   & 33.66/0.9213 & 29.77/0.8314 & 28.82/0.7976 & 27.14/0.8279 \\
& DRCN\cite{Kim_2016_DRCN}      & 1,774K & 17,974.3G & 33.82/0.9226 & 29.76/0.8311 & 28.80/0.7963 & 27.15/0.8276 \\
& CNF\cite{cnf2017}        & 337K   & 311.0G   & 33.74/0.9226 & 29.90/0.8322 & 28.82/0.7980 & -\\
& DRRN\cite{Tai-DRRN-2017}      & 297K   & 6,796.9G & 34.03/0.9244 & 29.96/0.8349 & 28.95/0.8004 & 27.53/0.8378 \\
& BTSRN\cite{fan2017balanced}      & 410K   & 176.2G   & 34.03/-\HSE  & 29.90/-\HSE  & 28.97/-\HSE & 27.75/-\HSE \\
& MemNet\cite{tai2017memnet}      & 677K   & 2,662.4G   & 34.09/0.9248 & 30.00/0.8350 & 28.96/0.8001 & 27.56/0.8376 \\
& SelNet\cite{choi2017deep}       & 1,159K & 120.0G   & 34.27/\textcolor{green}{0.9257}  & \textcolor{green}{30.30}/0.8399  & 28.97/0.8025 & - \\
& SRMDNF\cite{zhang2018learning}    &  -  &  -  & 34.12/0.9254 & 30.04/0.8382 & 28.97/0.8025 & 27.57/0.8398 \\
& CARN \cite{ahn2018fast}             & 1,592K & 118.8G   & \textcolor{green}{34.29}/0.9255 & 30.29/\textcolor{green}{0.8407} & \textcolor{green}{29.06}/\textcolor{green}{0.8034}	& \textcolor{green}{28.06}/\textcolor{green}{0.8493} \\
& CARN-M \cite{ahn2018fast}            & 412K   & 46.1G    & 33.99/0.9236 & 30.08/0.8367 & 28.91/0.8000	& 27.55/0.8385\\
& MAFFSRN (ours)      & 418K   & 34.2G   & \blue{34.32}/\blue{0.9269} & \blue{30.35}/\blue{0.8429} & \blue{29.09}/\blue{0.8052} & \blue{28.13}/\blue{0.8521} \\
& MAFFSRN-L (ours)      & 807K    & 68.5G    & \red{34.45}/\red{0.9277} & \red{30.40}/\red{0.8432} & \red{29.13}/\red{0.8061} & \red{28.26}/\red{0.8552} \Bstrut \\\hline\Tstrut
\multirow{13}{*}{4} &SRCNN\cite{dong2015image}
                          & 57K    & 52.7G    & 30.48/0.8628 & 27.49/0.7503 & 26.90/0.7101 & 24.52/0.7221 \\
& FSRCNN\cite{dong2016accelerating}  & 12K    & 4.6G     & 30.71/0.8657 & 27.59/0.7535 & 26.98/0.7150 & 24.62/0.7280 \\
& VDSR\cite{kim2016accurate}      & 665K   & 612.6G   & 31.35/0.8838 & 28.01/0.7674 & 27.29/0.7251 & 25.18/0.7524 \\
& DRCN\cite{Kim_2016_DRCN}      & 1,774K & 17,974.3G & 31.53/0.8854 & 28.02/0.7670 & 27.23/0.7233 & 25.14/0.7510 \\
& CNF\cite{cnf2017}        & 337K   & 311.0G   & 31.55/0.8856 & 28.15/0.7680 & 27.32/0.7253 & -\\
& LapSRN\cite{lai2017laplacian}  & 813K   & 149.4G   & 31.54/0.8850 & 28.19/0.7720 & 27.32/0.7280 & 25.21/0.7560 \\
& DRRN\cite{Tai-DRRN-2017}      & 297K   & 6,796.9G & 31.68/0.8888 & 28.21/0.7720 & 27.38/0.7284 & 25.44/0.7638\\
& BTSRN \cite{fan2017balanced}    & 410K   & 165.2G   & 31.85/-\HSE  & 28.20/-\HSE  & 27.47/-\HSE  & 25.74/-\HSE \\
& MemNet\cite{Tai-DRRN-2017}      & 677K   & 2,662.4G   & 31.74/0.8893 & 28.26/0.7723 & 27.40/0.7281 & 25.50/0.7630 \\
& SelNet\cite{choi2017deep}       & 1,417K & 83.1G    & 32.00/0.8931 & 28.49/0.7783 & 27.44/0.7325 & - \\
& SRDenseNet \cite{tong2017image} & 2,015K & 389.9G   & 32.02/0.8934 & 28.50/0.7782 & 27.53/0.7337 & 26.05/0.7819 \\
& SRMDNF\cite{zhang2018learning}     &  -  &    -     & 31.96/0.8925 & 28.35/0.7787 & 27.49/0.7337 & 25.68/0.7731 \\
& CARN \cite{ahn2018fast}             & 1,592K & 90.9G    & 32.13/0.8937 & \textcolor{green}{28.60}/0.7806 & \blue{27.58}/0.7349	& 26.07/0.7837 \\
& CARN-M  \cite{ahn2018fast}          & 412K   & 32.5G    & 31.92/0.8903 & 28.42/0.7762 & 27.44/0.7304 & 25.62/0.7694 \\
& CBPN-S\cite{zhu2019efficient}    & 592K   & 63.1G    & 31.93/0.8908 & 28.50/0.7785 & 27.50/0.7324 & 25.85/0.7772 \\
& CBPN\cite{zhu2019efficient}    & 1,197K & 97.9G    & \red{32.21}/\textcolor{green}{0.8944} & \red{28.63}/\blue{0.7813} & \blue{27.58}/\textcolor{green}{0.7356} & \blue{26.14}/\blue{0.7869} \\
& MAFFSRN (ours)      & 441K    & 19.3G    & \textcolor{green}{32.18}/\blue{0.8948} & 28.58/\textcolor{green}{0.7812} & \textcolor{green}{27.57}/\blue{0.7361} & \textcolor{green}{26.04}/\textcolor{green}{0.7848} \\
& MAFFSRN-L (ours)      & 830K    & 38.6G    & \blue{32.20}/\red{0.8953} & \blue{28.62}/\red{0.7822} & \red{27.59}/\red{0.7370} & \red{26.16}/\red{0.7887} \Bstrut\\\hline
\end{tabular}
}
\label{table:benchmark}

\end{table}

\subsection{Comparison with Existing Methods}

In this section, we present our quantitatively and qualitatively results and compare their performance with the state-of-the-art methods \cite{dong2015image,dong2016accelerating,kim2016accurate,Kim_2016_DRCN,cnf2017,lai2017laplacian,Tai-DRRN-2017,fan2017balanced,choi2017deep,chu2019fast,zhang2018learning,ahn2018fast,zhu2019efficient} on three up-scaling factors 2$\times$, 3$\times$ and 4$\times$. The quantitative results are shown in Table \ref{table:benchmark}. These also include the number of operations (Multi-Adds) and number of parameters to show the model complexity. Multi-Adds are estimated on 720p HR image. The results suggest that our lightweight MAFFSRN model achieves better performance than other methods on multiple datasets and scaling factors. 
Note that, our lightweight MAFFSRN model shows comparable performance to those models that consume 2$\times$ to 3$\times$ computing resources. 

Furthermore, to demonstrate the superiority of our model, we compare the performance (PSNR) and computational cost (Multi-Adds) of our models with the existing models in Figure \ref{fig:efficient_fig}. It is evident from figure that our methods outperform the existing networks in both complexity and PSNR. It is worth to note that our MAFFSRN model even consists of fewer Multi-Adds than SRCNN \cite{dong2015image} which is a shallow neural network with 3-layers. 

\begin{figure}[!t]
	\newlength\fsdttwofigBD

	\setlength{\fsdttwofigBD}{-1.5mm}
	\scriptsize
	\centering
	\begin{tabular}{cc}
		
		\begin{adjustbox}{valign=t}
		\tiny
			\begin{tabular}{c}
				\includegraphics[width=0.229\textwidth]{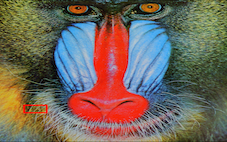}
				\\
				Set14 ($2\times$):
				\\
				baboon
			\end{tabular}
		\end{adjustbox}
		\hspace{-2.3mm}
		\begin{adjustbox}{valign=t}
		\tiny
			\begin{tabular}{cccccc}
				\includegraphics[width=\widthscalefive \textwidth]{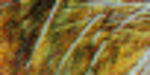} \hspace{\fsdttwofigBD} &
				\includegraphics[width=\widthscalefive \textwidth]{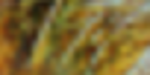} \hspace{\fsdttwofigBD} &
				\includegraphics[width=\widthscalefive \textwidth]{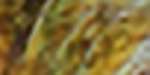} \hspace{\fsdttwofigBD} &
				\includegraphics[width=\widthscalefive \textwidth]{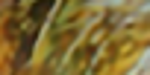} \hspace{\fsdttwofigBD} &
				\includegraphics[width=\widthscalefive \textwidth]{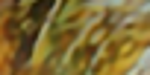} 
				\\
				HR \hspace{\fsdttwofigBD} &
				Bicubic \hspace{\fsdttwofigBD} &
				SRCNN~\cite{dong2016image} \hspace{\fsdttwofigBD} &
				VDSR~\cite{kim2016accurate} &
				LapSRN~\cite{lai2017laplacian} &
				\\
				PSNR/SSIM \hspace{\fsdttwofigBD} &
				24.86/0.6954 \hspace{\fsdttwofigBD} &
			    25.74/0.7669 \hspace{\fsdttwofigBD} &
				25.95/0.7792 \hspace{\fsdttwofigBD} &
				25.89/0.7781 \hspace{\fsdttwofigBD} &
				\\
				\includegraphics[width=\widthscalefive \textwidth]{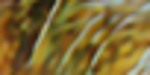} \hspace{\fsdttwofigBD} &
				\includegraphics[width=\widthscalefive \textwidth]{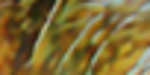} \hspace{\fsdttwofigBD} &
				\includegraphics[width=\widthscalefive \textwidth]{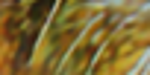} \hspace{\fsdttwofigBD} &
				\includegraphics[width=\widthscalefive \textwidth]{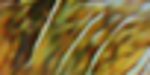} \hspace{\fsdttwofigBD} &
				\includegraphics[width=\widthscalefive \textwidth]{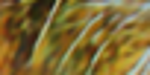}  
				\\ 
				SRMDNF~\cite{zhang2018learning} \hspace{\fsdttwofigBD} &
				CARN-M~\cite{ahn2018fast} \hspace{\fsdttwofigBD} &
				CARN~\cite{ahn2018fast} \hspace{\fsdttwofigBD} &
				\textbf{MAFFSRN}~  \hspace{\fsdttwofigBD} &
				\textbf{MAFFSRN-L}
				\\
				 26.12/0.7859\hspace{\fsdttwofigBD} &
				 26.18/0.7883\hspace{\fsdttwofigBD} &
				 26.34/0.7953\hspace{\fsdttwofigBD} &
				 \textbf{26.40}/\textbf{0.7967}\hspace{\fsdttwofigBD} &
				\textbf{26.49}/\textbf{0.8000} \hspace{\fsdttwofigBD} 
				\\
			\end{tabular}
		\end{adjustbox}
		\vspace{3mm}
		\\
		\begin{adjustbox}{valign=t}
		\tiny
			\begin{tabular}{c}
				\includegraphics[width=0.229\textwidth]{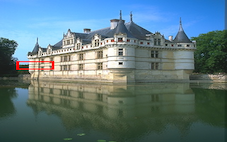}
				\\
				 B100 ($2\times$):
				\\
				102061
				
			\end{tabular}
		\end{adjustbox}
		\hspace{-2.0mm}
		\begin{adjustbox}{valign=t}
		\tiny
			\begin{tabular}{cccccc}
				\includegraphics[width=\widthscalefive \textwidth]{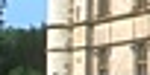} \hspace{\fsdttwofigBD} &
				\includegraphics[width=\widthscalefive \textwidth]{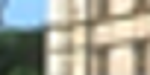} \hspace{\fsdttwofigBD} &
				\includegraphics[width=\widthscalefive \textwidth]{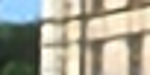} \hspace{\fsdttwofigBD} &
				\includegraphics[width=\widthscalefive \textwidth]{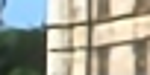} \hspace{\fsdttwofigBD} &
				\includegraphics[width=\widthscalefive \textwidth]{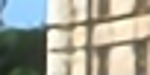} 
				\\
				HR \hspace{\fsdttwofigBD} &
				Bicubic \hspace{\fsdttwofigBD} &
				SRCNN~\cite{dong2016image} \hspace{\fsdttwofigBD} &
				VDSR~\cite{kim2016accurate} &
				LapSRN~\cite{lai2017laplacian} &
				\\
				PSNR/SSIM \hspace{\fsdttwofigBD} &
				 28.09/0.8703\hspace{\fsdttwofigBD} &
				 29.45/0.9019\hspace{\fsdttwofigBD} &
				 30.09/0.9127\hspace{\fsdttwofigBD} &
				 29.90/0.9113\hspace{\fsdttwofigBD} &
				\\
				\includegraphics[width=\widthscalefive \textwidth]{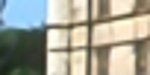} \hspace{\fsdttwofigBD} &
				\includegraphics[width=\widthscalefive \textwidth]{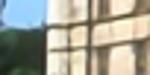} \hspace{\fsdttwofigBD} &
				\includegraphics[width=\widthscalefive \textwidth]{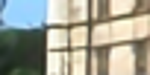} \hspace{\fsdttwofigBD} &
				\includegraphics[width=\widthscalefive \textwidth]{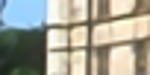} \hspace{\fsdttwofigBD} &
				\includegraphics[width=\widthscalefive \textwidth]{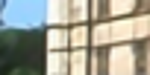}  
				\\ 
				SRMDNF~\cite{zhang2018learning} \hspace{\fsdttwofigBD} &
				CARN-M~\cite{ahn2018fast} \hspace{\fsdttwofigBD} &
				CARN~\cite{ahn2018fast} \hspace{\fsdttwofigBD} &
				\textbf{MAFFSRN}~  \hspace{\fsdttwofigBD} &
			\textbf{MAFFSRN-L} 
				\\
				30.35/0.9160 \hspace{\fsdttwofigBD} &
				30.37/0.9161\hspace{\fsdttwofigBD} &
				30.38/0.9158\hspace{\fsdttwofigBD} &
				\textbf{30.46}/\textbf{0.9172}\hspace{\fsdttwofigBD} &
				\textbf{30.58}/\textbf{0.9185} \hspace{\fsdttwofigBD} 
				\\
				
			\end{tabular}
		\end{adjustbox}
		\vspace{1mm}
		\\
		\begin{adjustbox}{valign=t}
		\tiny
			\begin{tabular}{c}
				\includegraphics[width=0.229\textwidth]{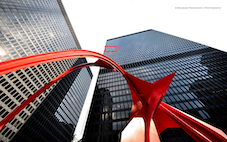}
				\\
				 Urban100 ($2\times$):
				\\
				img\_062
				
			\end{tabular}
		\end{adjustbox}
		\hspace{-2.0mm}
		\begin{adjustbox}{valign=t}
		\tiny
			\begin{tabular}{cccccc}
				\includegraphics[width=\widthscalefive \textwidth]{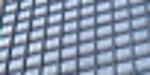} \hspace{\fsdttwofigBD} &
				\includegraphics[width=\widthscalefive \textwidth]{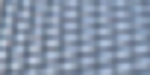} \hspace{\fsdttwofigBD} &
				\includegraphics[width=\widthscalefive \textwidth]{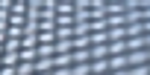} \hspace{\fsdttwofigBD} &
				\includegraphics[width=\widthscalefive \textwidth]{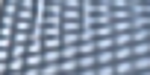} \hspace{\fsdttwofigBD} &
				\includegraphics[width=\widthscalefive \textwidth]{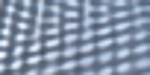} 
				\\
				HR \hspace{\fsdttwofigBD} &
				Bicubic \hspace{\fsdttwofigBD} &
				SRCNN~\cite{dong2016image} \hspace{\fsdttwofigBD} &
				VDSR~\cite{kim2016accurate} &
				LapSRN~\cite{lai2017laplacian} &
				\\
				PSNR/SSIM \hspace{\fsdttwofigBD} & 
				23.44/0.8527 \hspace{\fsdttwofigBD} &
				26.04/0.9217 \hspace{\fsdttwofigBD} &
				27.12/0.9442 \hspace{\fsdttwofigBD} &
				26.68/0.9395 \hspace{\fsdttwofigBD} &
				\\
				\includegraphics[width=\widthscalefive \textwidth]{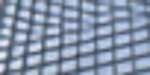} \hspace{\fsdttwofigBD} &
				\includegraphics[width=\widthscalefive \textwidth]{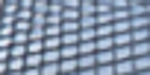} \hspace{\fsdttwofigBD} &
				\includegraphics[width=\widthscalefive \textwidth]{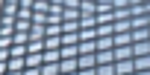} \hspace{\fsdttwofigBD} &
				\includegraphics[width=\widthscalefive \textwidth]{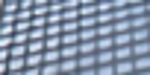} \hspace{\fsdttwofigBD} &
				\includegraphics[width=\widthscalefive \textwidth]{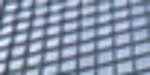}  
				\\ 
				SRMDNF~\cite{zhang2018learning} \hspace{\fsdttwofigBD} &
				CARN-M~ \cite{ahn2018fast} \hspace{\fsdttwofigBD} &
				CARN~\cite{ahn2018fast} \hspace{\fsdttwofigBD} &
				\textbf{MAFFSRN}~  \hspace{\fsdttwofigBD} &
				\textbf{MAFFSRN-L} 
				\\
				27.95/0.9549 \hspace{\fsdttwofigBD} &
				 27.83/0.9538 \hspace{\fsdttwofigBD} &
				29.19/0.9637 \hspace{\fsdttwofigBD} &
				\textbf{29.65}/\textbf{0.9647} \hspace{\fsdttwofigBD} &
				\textbf{30.09}/\textbf{0.9684} \hspace{\fsdttwofigBD} 
				\\
			\end{tabular}
		\end{adjustbox}
		\vspace{-3mm}
		\\
	\end{tabular}
	\caption{
		Visual comparison for $2\times$ SR with other models on Set14, B100, Urban10 dataset. The best results are \textbf{highlighted}
	}
	\label{fig:result_BI2x}

\end{figure}

\begin{figure}[t]
	\setlength{\fsdttwofigBD}{-1.5mm}
	\scriptsize
	\centering
	\begin{tabular}{cc}
		
		\begin{adjustbox}{valign=t}
		\tiny
			\begin{tabular}{c}
				\includegraphics[width=0.229\textwidth]{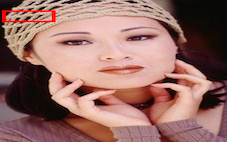}
				\\
				Set5 ($4\times$):
				\\
				woman
			\end{tabular}
		\end{adjustbox}
		\hspace{-2.3mm}
		\begin{adjustbox}{valign=t}
		\tiny
			\begin{tabular}{cccccc}
				\includegraphics[width=\widthscalefive \textwidth]{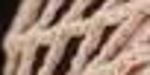} \hspace{\fsdttwofigBD} &
				\includegraphics[width=\widthscalefive \textwidth]{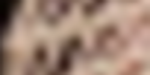} \hspace{\fsdttwofigBD} &
				\includegraphics[width=\widthscalefive \textwidth]{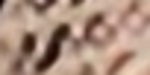} \hspace{\fsdttwofigBD} &
				\includegraphics[width=\widthscalefive \textwidth]{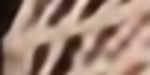} \hspace{\fsdttwofigBD} &
				\includegraphics[width=\widthscalefive \textwidth]{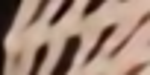} 
				\\
				HR \hspace{\fsdttwofigBD} &
				Bicubic \hspace{\fsdttwofigBD} &
				SRCNN~\cite{dong2016image} \hspace{\fsdttwofigBD} & 
				VDSR~\cite{kim2016accurate} & 
				LapSRN~\cite{lai2017laplacian} & 
				\\
				PSNR/SSIM \hspace{\fsdttwofigBD} &
				26.47/0.8325 \hspace{\fsdttwofigBD} &
				28.88/0.8837 \hspace{\fsdttwofigBD} &
				29.81/0.9049 \hspace{\fsdttwofigBD} &
				30.32/0.9091 \hspace{\fsdttwofigBD} &
				 
				\\
				\includegraphics[width=\widthscalefive \textwidth]{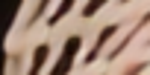} \hspace{\fsdttwofigBD} &
				\includegraphics[width=\widthscalefive \textwidth]{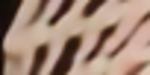} \hspace{\fsdttwofigBD} &
				\includegraphics[width=\widthscalefive \textwidth]{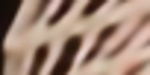} \hspace{\fsdttwofigBD} &
				\includegraphics[width=\widthscalefive \textwidth]{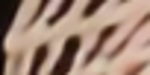} \hspace{\fsdttwofigBD} &
				\includegraphics[width=\widthscalefive \textwidth]{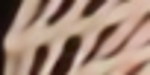}  
				\\ 
				SRMDNF~\cite{zhang2018learning} \hspace{\fsdttwofigBD} &
				CARN-M~\cite{ahn2018fast} \hspace{\fsdttwofigBD} &
				CARN~\cite{ahn2018fast} \hspace{\fsdttwofigBD} &
				\textbf{MAFFSRN}~  \hspace{\fsdttwofigBD} &
				\textbf{MAFFSRN-L}
				\\
				30.45/0.9126 \hspace{\fsdttwofigBD} &
				30.54/0.9110 \hspace{\fsdttwofigBD} &
				30.50/0.9133 \hspace{\fsdttwofigBD} &
				\textbf{30.59}/\textbf{0.9146} \hspace{\fsdttwofigBD} &
				\textbf{30.73}/\textbf{0.9153} \hspace{\fsdttwofigBD} 
				\\
			\end{tabular}
		\end{adjustbox}
		\vspace{0.2mm}
		\\
		\begin{adjustbox}{valign=t}
		\tiny
			\begin{tabular}{c}
				\includegraphics[width=0.229\textwidth]{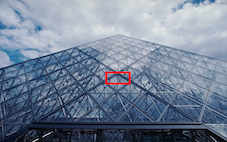}
				\\
				 B100 ($4\times$):
				\\
				223061
				
			\end{tabular}
		\end{adjustbox}
		\hspace{-2.3mm}
		\begin{adjustbox}{valign=t}
		\tiny
			\begin{tabular}{cccccc}
				\includegraphics[width=\widthscalefive \textwidth]{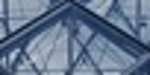} \hspace{\fsdttwofigBD} &
				\includegraphics[width=\widthscalefive \textwidth]{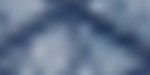} \hspace{\fsdttwofigBD} &
				\includegraphics[width=\widthscalefive \textwidth]{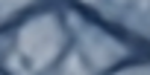} \hspace{\fsdttwofigBD} &
				\includegraphics[width=\widthscalefive \textwidth]{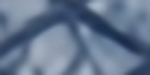} \hspace{\fsdttwofigBD} &
				\includegraphics[width=\widthscalefive \textwidth]{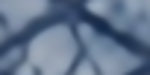} 
				\\
				HR \hspace{\fsdttwofigBD} &
				Bicubic \hspace{\fsdttwofigBD} &
				SRCNN~\cite{dong2016image} \hspace{\fsdttwofigBD} &
				VDSR~\cite{kim2016accurate} &
				LapSRN~\cite{lai2017laplacian} &
				\\
				PSNR/SSIM \hspace{\fsdttwofigBD} &
				23.86/0.5909 \hspace{\fsdttwofigBD} &
				24.32/0.6424 \hspace{\fsdttwofigBD} &
				24.46/0.6538 \hspace{\fsdttwofigBD} &
				24.51/0.6573 \hspace{\fsdttwofigBD} &
				\\
				\includegraphics[width=\widthscalefive \textwidth]{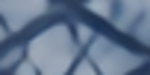} \hspace{\fsdttwofigBD} &
				\includegraphics[width=\widthscalefive \textwidth]{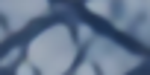} \hspace{\fsdttwofigBD} &
				\includegraphics[width=\widthscalefive \textwidth]{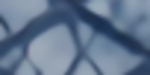} \hspace{\fsdttwofigBD} &
				\includegraphics[width=\widthscalefive \textwidth]{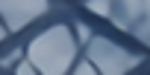} \hspace{\fsdttwofigBD} &
				\includegraphics[width=\widthscalefive \textwidth]{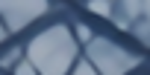}  
				\\ 
				SRMDNF~\cite{zhang2018learning} \hspace{\fsdttwofigBD} &
				CARN-M~\cite{ahn2018fast} \hspace{\fsdttwofigBD} &
				CARN~\cite{ahn2018fast} \hspace{\fsdttwofigBD} &
				\textbf{MAFFSRN}~  \hspace{\fsdttwofigBD} &
				\textbf{MAFFSRN-L} 
				\\
				24.69/0.6729 \hspace{\fsdttwofigBD} &
				24.64/0.6642 \hspace{\fsdttwofigBD} &
				24.75/0.6752 \hspace{\fsdttwofigBD} &
				\textbf{24.81}/\textbf{0.6850} \hspace{\fsdttwofigBD} &
				\textbf{24.84}/\textbf{0.6899} \hspace{\fsdttwofigBD}
				
				\textbf{}/\textbf{} \hspace{\fsdttwofigBD} 
				\\
			\end{tabular}
		\end{adjustbox}
		\vspace{0.5mm}
		\\
		\begin{adjustbox}{valign=t}
		\tiny
			\begin{tabular}{c}
				\includegraphics[width=0.229\textwidth]{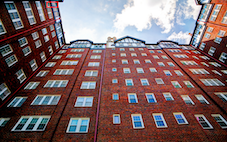}
				\\
				 Urban100 ($4\times$):
				\\
				img\_034
				
			\end{tabular}
		\end{adjustbox}
		\hspace{-2.3mm}
		\begin{adjustbox}{valign=t}
		\tiny
			\begin{tabular}{cccccc}
				\includegraphics[width=\widthscalefive \textwidth]{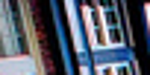} \hspace{\fsdttwofigBD} &
				\includegraphics[width=\widthscalefive \textwidth]{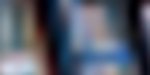} \hspace{\fsdttwofigBD} &
				\includegraphics[width=\widthscalefive \textwidth]{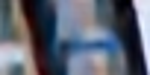} \hspace{\fsdttwofigBD} &
				\includegraphics[width=\widthscalefive \textwidth]{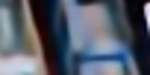} \hspace{\fsdttwofigBD} &
				\includegraphics[width=\widthscalefive \textwidth]{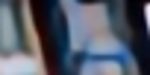} 
				\\
				HR \hspace{\fsdttwofigBD} &
				Bicubic \hspace{\fsdttwofigBD} &
				SRCNN~\cite{dong2016image} \hspace{\fsdttwofigBD} &
				VDSR~\cite{kim2016accurate} &
				LapSRN~\cite{lai2017laplacian} &
				\\
				PSNR/SSIM \hspace{\fsdttwofigBD} &
				21.41/0.4799 \hspace{\fsdttwofigBD} &
				22.32/0.5448 \hspace{\fsdttwofigBD} &
				22.62/0.5648 \hspace{\fsdttwofigBD} &
				22.65/0.5648 \hspace{\fsdttwofigBD} &
				 
				\\
				\includegraphics[width=\widthscalefive \textwidth]{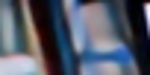} \hspace{\fsdttwofigBD} &
				\includegraphics[width=\widthscalefive \textwidth]{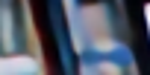} \hspace{\fsdttwofigBD} &
				\includegraphics[width=\widthscalefive \textwidth]{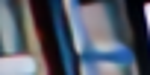} \hspace{\fsdttwofigBD} &
				\includegraphics[width=\widthscalefive \textwidth]{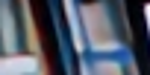} \hspace{\fsdttwofigBD} &
				\includegraphics[width=\widthscalefive \textwidth]{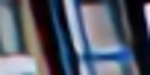}  
				\\ 
				SRMDNF~\cite{zhang2018learning} \hspace{\fsdttwofigBD} &
				CARN-M~\cite{ahn2018fast} \hspace{\fsdttwofigBD} &
				CARN~\cite{ahn2018fast} \hspace{\fsdttwofigBD} &
			    \textbf{MAFFSRN}~  \hspace{\fsdttwofigBD} &
				\textbf{MAFFSRN-L} 
				\\
				22.98/0.5817 \hspace{\fsdttwofigBD} &
				22.88/0.5764 \hspace{\fsdttwofigBD} &
				23.09/0.5861 \hspace{\fsdttwofigBD} &
				\textbf{23.11}/\textbf{0.5858} \hspace{\fsdttwofigBD} &
				 \textbf{23.13}/\textbf{0.5870} \hspace{\fsdttwofigBD} 
				\\
			\end{tabular}
		\end{adjustbox}
		\vspace{-3mm}
		\\
	\end{tabular}
	\caption{
		Visual comparison for $4\times$ SR with other models on Set5, B100, Urban10 dataset. The best results are \textbf{highlighted}
	}
	\label{fig:result_BI4x}

\end{figure}

We present our qualitative results in Figure \ref{fig:result_BI2x} and Figure \ref{fig:result_BI4x}. In Figure \ref{fig:result_BI2x}, it can be seen from output results that the hairs of 'baboon' moustache are accurately reconstructed whereas other methods show blurry results. The similar effects can be seen in other images of Figure \ref{fig:result_BI2x} where our methods demonstrate superior results. The results also include PSNR to show the qualitative results. Furthermore, our method continues to show improved results in even larger scale $4\times$. Overall, our methods have shown improved results as compared to existing methods.






\begin{figure}[h]
  \includegraphics[width=1.0\linewidth]{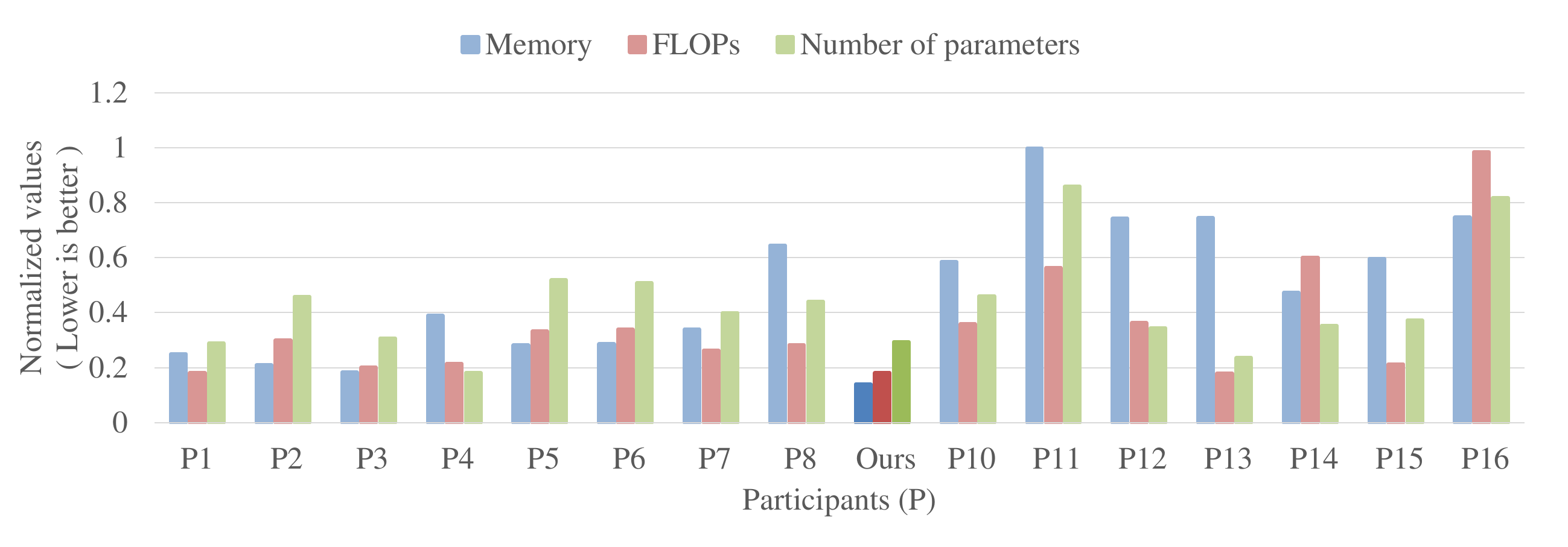} 
  \caption{Computational cost of all participants. Normalized values are shown on the x-axis whereas participants are shown on the y-axis. The results are shown in dark colors (ours) and light colors (other participants)}
  \label{fig:AIMbar}
\end{figure}

\section{AIM2020 Efficient SR Challenge}

\begin{table}[!h]
\centering
\caption{Performance comparison of each entry in the AIM2020 efficient SR challenge. The number in the parenthesis denotes the rank}

\begin{tabular}{P{0.2\textwidth}P{0.13\textwidth}P{0.13\textwidth}P{0.13\textwidth}}

\hline
Method& Memory& FLOPs & Parameters \\ 
     &  [MB]& [G] & [M] \\ \hline
MAFFSRN        &112(1)	    &27.11(3)	&0.441(4)   \\ 
Participant 1 & 146(2)	&30.06(4)	&0.461(5)   \\ 
Participant 2 & 168(3)	&44.98(9)	&0.687(11)  \\ 
Participant 3 & 200(4)	&27.10(2)	&0.433(3)   \\ 
Participant 4 & 225(5)	&49.67(10)	&0.777(14)  \\ 
Participant 5 & 229(6)	&50.85(11)	&0.761(13)  \\
\hline
Baseline   & 610 &    166.36 & 1.517     \\ 
\hline
\end{tabular}
\label{table:AIMtable}
\end{table}

Our model is developed to participate in the AIM 2020 efficient SR challenge \cite{zhang2020aim}.  This competition targets to develop a practicable SR method that can be utilized in a constrained environment. The aim was to maintain the PSNR of MSResNet  \cite{wang2018esrgan} on DIV2K \cite{timofte2017ntire} validation set while decreasing its computational cost. 
We submitted our MAFFSRN model to this challenge and won 1st in Memory computations, 3rd in FLOPs, 4th in number of parameters. 

We present the results in Figure \ref{fig:AIMbar} and Table \ref{table:AIMtable}. Figure  \ref{fig:AIMbar} shows the normalized scores on the y-axis and final participants on the x-axis. It also indicates that our proposed MAFFSRN model is a lightweight model among all participants. Similarly, we show the performance of top participants in Table \ref{table:AIMtable}  sorted by memory computations. Note that memory computations are tested with Pytorch code \textit{torch.cuda.max\_memory\_allocated()} and FLOPs are calculated with an input image $256\times256$.



\section{Limitation and Future Work}

In spite of having reduced computational cost (memory consumption, FLOPs, number of parameters), the runtime of the proposed method on the 100 validation images of DIV2K dataset \cite{timofte2017ntire} is  0.104 sec per image\footnote{it is reported in \cite{zhang2020aim} }. The estimated runtime is averaged over  100 images. We assume it is a consequence of more layers in our networks than other efficient architectures. Nevertheless, the proposed method is ultra lightweight, and its memory-efficient modules can assist future researchers in the advancement of efficient SR architectures that have lower runtime and reduced memory consumption.

\section{Conclusion}

This work introduces a lightweight SR method for a constrained environment called MAFFSRN. We show with the several quantitative and qualitative experiments that MAFFSRN outperforms other existing lightweight models in terms of both performance and computational cost. Further, we present ablation studies to show the contributions of each proposed module.

\section*{Acknowledgement}
This research was supported by Basic Science Research Program through the National Research Foundation of Korea (NRF) funded by the Ministry of  Science, ICT \& Future Planning (2018R1C1B3008159).
Also, this research was a result of a study on the "HPC Support" Project, supported by the ‘Ministry of Science and ICT’ and NIPA.

\bibliographystyle{splncs04}
\bibliography{egbib}

\begin{thebibliography}{10}
\providecommand{\url}[1]{\texttt{#1}}
\providecommand{\urlprefix}{URL }
\providecommand{\doi}[1]{https://doi.org/#1}

\bibitem{agustsson2017ntire}
Agustsson, E., Timofte, R.: Ntire 2017 challenge on single image
  super-resolution: Dataset and study. In: Proc. Computer Vision and Pattern
  Recognition~(CVPR) Workshops. pp. 126--135 (2017)

\bibitem{ahn2018fast}
Ahn, N., Kang, B., Sohn, K.A.: Fast, accurate, and lightweight super-resolution
  with cascading residual network. In: Proc. European Conference on Computer
  Vision~(ECCV). pp. 252--268 (2018)

\bibitem{bevilacqua2012low}
Bevilacqua, M., Roumy, A., Guillemot, C., Alberi{-}Morel, M.: Low-complexity
  single-image super-resolution based on nonnegative neighbor embedding. In:
  Proc. British Machine Vision Conference~(BMVC). pp. 1--10 (2012)

\bibitem{cai2020learning}
Cai, Y., Wang, Z., Luo, Z., Yin, B., Du, A., Wang, H., Zhou, X., Zhou, E.,
  Zhang, X., Sun, J.: Learning delicate local representations for multi-person
  pose estimation. arXiv preprint arXiv:2003.04030  (2020)

\bibitem{choi2017deep}
Choi, J.S., Kim, M.: A deep convolutional neural network with selection units
  for super-resolution. In: Proc. Computer Vision and Pattern
  Recognition~(CVPR) Workshops. pp. 154--160 (2017)

\bibitem{chollet2017xception}
Chollet, F.: Xception: Deep learning with depthwise separable convolutions. In:
  Proc. Computer Vision and Pattern Recognition~(CVPR). pp. 1251--1258 (2017)

\bibitem{chu2019fast}
Chu, X., Zhang, B., Ma, H., Xu, R., Li, J., Li, Q.: Fast, accurate and
  lightweight super-resolution with neural architecture search. arXiv preprint
  arXiv:1901.07261  (2019)

\bibitem{dong2015image}
Dong, C., Loy, C.C., He, K., Tang, X.: Image super-resolution using deep
  convolutional networks. Pattern Analysis and Machine Intelligence~(PAMI)
  \textbf{38}(2),  295--307 (2015)

\bibitem{dong2016image}
Dong, C., Loy, C.C., He, K., Tang, X.: Image super-resolution using deep
  convolutional networks. Pattern Analysis and Machine Intelligence~(PAMI)
  \textbf{38}(2),  295--307 (2016)

\bibitem{dong2016accelerating}
Dong, C., Loy, C.C., Tang, X.: Accelerating the super-resolution convolutional
  neural network. In: Proc. European Conference on Computer Vision~(ECCV). pp.
  391--407. Springer (2016)

\bibitem{fan2017balanced}
Fan, Y., Shi, H., Yu, J., Liu, D., Han, W., Yu, H., Wang, Z., Wang, X., Huang,
  T.S.: Balanced two-stage residual networks for image super-resolution. In:
  Proc. Computer Vision and Pattern Recognition~(CVPR) Workshops. pp. 161--168
  (2017)

\bibitem{han2015deep}
Han, S., Mao, H., Dally, W.J.: Deep compression: Compressing deep neural
  networks with pruning, trained quantization and huffman coding. arXiv
  preprint arXiv:1510.00149  (2015)

\bibitem{haris2018deep}
Haris, M., Shakhnarovich, G., Ukita, N.: Deep back-projection networks for
  super-resolution. In: Proc. Computer Vision and Pattern Recognition~(CVPR)
  (2018)

\bibitem{he2016deep}
He, K., Zhang, X., Ren, S., Sun, J.: Deep residual learning for image
  recognition. In: Proc. Computer Vision and Pattern Recognition~(CVPR). pp.
  770--778 (2016)

\bibitem{heo2020slowing}
Heo, B., Chun, S., Oh, S.J., Han, D., Yun, S., Uh, Y., Ha, J.W.: Slowing down
  the weight norm increase in momentum-based optimizers. arXiv preprint
  arXiv:2006.08217  (2020)

\bibitem{hinton2015distilling}
Hinton, G., Vinyals, O., Dean, J.: Distilling the knowledge in a neural
  network. arXiv preprint arXiv:1503.02531  (2015)

\bibitem{hu2018squeeze}
Hu, J., Shen, L., Sun, G.: Squeeze-and-excitation networks. In: Proc. Computer
  Vision and Pattern Recognition~(CVPR). pp. 7132--7141 (2018)

\bibitem{huang2017densely}
Huang, G., Liu, Z., Van Der~Maaten, L., Weinberger, K.Q.: Densely connected
  convolutional networks. In: Proc. Computer Vision and Pattern
  Recognition~(CVPR). pp. 4700--4708 (2017)

\bibitem{huang2015single}
Huang, J.B., Singh, A., Ahuja, N.: Single image super-resolution from
  transformed self-exemplars. In: Proc. Computer Vision and Pattern
  Recognition~(CVPR). pp. 5197--5206 (2015)

\bibitem{ioffe2015batch}
Ioffe, S., Szegedy, C.: Batch normalization: Accelerating deep network training
  by reducing internal covariate shift. arXiv preprint arXiv:1502.03167  (2015)

\bibitem{kim2016accurate}
Kim, J., Lee, J.K., Lee, K.M.: Accurate image super-resolution using very deep
  convolutional networks. In: cvpr (June 2016)

\bibitem{Kim_2016_DRCN}
Kim, J., Lee, J.K., Lee, K.M.: Deeply-recursive convolutional network for image
  super-resolution. In: cvpr (June 2016)

\bibitem{kim2018ram}
Kim, J.H., Choi, J.H., Cheon, M., Lee, J.S.: Ram: Residual attention module for
  single image super-resolution. arXiv preprint arXiv:1811.12043  (2018)

\bibitem{kingma2014adam}
Kingma, D.P., Ba, J.: Adam: A method for stochastic optimization. arXiv
  preprint arXiv:1412.6980  (2014)

\bibitem{lai2017laplacian}
Lai, W.S., Huang, J.B., Ahuja, N., Yang, M.H.: Deep laplacian pyramid networks
  for fast and accurate super-resolution. In: cvpr (2017)

\bibitem{ledig2017photo}
Ledig, C., Theis, L., Husz{\'a}r, F., Caballero, J., Cunningham, A., Acosta,
  A., Aitken, A., Tejani, A., Totz, J., Wang, Z., et~al.: Photo-realistic
  single image super-resolution using a generative adversarial network. In:
  Proc. Computer Vision and Pattern Recognition~(CVPR). pp. 4681--4690 (2017)

\bibitem{li2018multi}
Li, J., Fang, F., Mei, K., Zhang, G.: Multi-scale residual network for image
  super-resolution. In: Proc. European Conference on Computer Vision~(ECCV).
  pp. 517--532 (2018)

\bibitem{Li_2018_ECCV}
Li, J., Fang, F., Mei, K., Zhang, G.: Multi-scale residual network for image
  super-resolution. In: Proc. European Conference on Computer Vision~(ECCV)
  (September 2018)

\bibitem{lim2017enhanced}
Lim, B., Son, S., Kim, H., Nah, S., Lee, K.M.: Enhanced deep residual networks
  for single image super-resolution. In: Proc. Computer Vision and Pattern
  Recognition~(CVPR) Workshops (July 2017)

\bibitem{liu2020residual}
Liu, J., Zhang, W., Tang, Y., Tang, J., Wu, G.: Residual feature aggregation
  network for image super-resolution. In: Proc. Computer Vision and Pattern
  Recognition~(CVPR). pp. 2359--2368 (2020)

\bibitem{martin2001database}
Martin, D., Fowlkes, C., Tal, D., Malik, J.: A database of human segmented
  natural images and its application to evaluating segmentation algorithms and
  measuring ecological statistics. In: Proc. International Conference on
  Computer Vision~(ICCV). vol.~2, pp. 416--423. IEEE (2001)

\bibitem{muqeet2019hybrid}
Muqeet, A., Iqbal, M.T.B., Bae, S.H.: Hybrid residual attention network for
  single image super resolution. arXiv preprint arXiv:1907.05514  (2019)

\bibitem{cnf2017}
Ren, H., El-Khamy, M., Lee, J.: Image super resolution based on fusing multiple
  convolution neural networks. In: Proc. Computer Vision and Pattern
  Recognition~(CVPR) Workshops. pp. 54--61 (2017)

\bibitem{shi2016real}
Shi, W., Caballero, J., Husz{\'a}r, F., Totz, J., Aitken, A.P., Bishop, R.,
  Rueckert, D., Wang, Z.: Real-time single image and video super-resolution
  using an efficient sub-pixel convolutional neural network. In: Proc. Computer
  Vision and Pattern Recognition~(CVPR). pp. 1874--1883 (2016)

\bibitem{Tai-DRRN-2017}
Tai, Y., Yang, J., Liu, X.: Image super-resolution via deep recursive residual
  network. In: Proc. Computer Vision and Pattern Recognition~(CVPR) (2017)

\bibitem{tai2017memnet}
Tai, Y., Yang, J., Liu, X., Xu, C.: Memnet: A persistent memory network for
  image restoration. In: Proc. International Conference on Computer
  Vision~(ICCV) (2017)

\bibitem{timofte2017ntire}
Timofte, R., Agustsson, E., Van~Gool, L., Yang, M.H., Zhang, L.: Ntire 2017
  challenge on single image super-resolution: Methods and results. In: Proc.
  Computer Vision and Pattern Recognition~(CVPR) Workshops. pp. 114--125 (2017)

\bibitem{tong2017image}
Tong, T., Li, G., Liu, X., Gao, Q.: Image super-resolution using dense skip
  connections. In: Proc. International Conference on Computer Vision~(ICCV).
  pp. 4799--4807 (2017)

\bibitem{urban2016deep}
Urban, G., Geras, K.J., Kahou, S.E., Aslan, O., Wang, S., Caruana, R., Mohamed,
  A., Philipose, M., Richardson, M.: Do deep convolutional nets really need to
  be deep and convolutional? arXiv preprint arXiv:1603.05691  (2016)

\bibitem{wang2019lightweight}
Wang, C., Li, Z., Shi, J.: Lightweight image super-resolution with adaptive
  weighted learning network. arXiv preprint arXiv:1904.02358  (2019)

\bibitem{wang2018esrgan}
Wang, X., Yu, K., Wu, S., Gu, J., Liu, Y., Dong, C., Qiao, Y., Change~Loy, C.:
  Esrgan: Enhanced super-resolution generative adversarial networks. In: Proc.
  European Conference on Computer Vision~(ECCV). pp.~0--0 (2018)

\bibitem{wang2004image}
Wang, Z., Bovik, A.C., Sheikh, H.R., Simoncelli, E.P., et~al.: Image quality
  assessment: from error visibility to structural similarity. IEEE transactions
  on image processing  \textbf{13}(4),  600--612 (2004)

\bibitem{woo2018cbam}
Woo, S., Park, J., Lee, J.Y., So~Kweon, I.: Cbam: Convolutional block attention
  module. In: Proc. European Conference on Computer Vision~(ECCV). pp. 3--19
  (2018)

\bibitem{yu2017dilated}
Yu, F., Koltun, V., Funkhouser, T.: Dilated residual networks. In: Proc.
  Computer Vision and Pattern Recognition~(CVPR). pp. 472--480 (2017)

\bibitem{zeyde2010single}
Zeyde, R., Elad, M., Protter, M.: On single image scale-up using
  sparse-representations. In: International conference on curves and surfaces.
  pp. 711--730. Springer (2010)

\bibitem{zhang2020aim}
Zhang, K., Danelljan, M., Li, Y., Timofte, R., et~al.: Aim 2020 challenge on
  efficient super-resolution: Methods and results. In: European Conference on
  Computer Vision Workshops (2020)

\bibitem{zhang2020deep}
Zhang, K., Gool, L.V., Timofte, R.: Deep unfolding network for image
  super-resolution. In: Proc. Computer Vision and Pattern Recognition~(CVPR).
  pp. 3217--3226 (2020)

\bibitem{zhang2018learning}
Zhang, K., Zuo, W., Zhang, L.: Learning a single convolutional super-resolution
  network for multiple degradations. In: Proc. Computer Vision and Pattern
  Recognition~(CVPR). pp. 3262--3271 (2018)

\bibitem{zhang2018shufflenet}
Zhang, X., Zhou, X., Lin, M., Sun, J.: Shufflenet: An extremely efficient
  convolutional neural network for mobile devices. In: Proc. Computer Vision
  and Pattern Recognition~(CVPR). pp. 6848--6856 (2018)

\bibitem{zhang2018image}
Zhang, Y., Li, K., Li, K., Wang, L., Zhong, B., Fu, Y.: Image super-resolution
  using very deep residual channel attention networks. In: Proc. European
  Conference on Computer Vision~(ECCV). pp. 286--301 (2018)

\bibitem{zhang2018residual}
Zhang, Y., Tian, Y., Kong, Y., Zhong, B., Fu, Y.: Residual dense network for
  image super-resolution. In: Proc. Computer Vision and Pattern
  Recognition~(CVPR) (2018)

\bibitem{zhu2019efficient}
Zhu, F., Zhao, Q.: Efficient single image super-resolution via hybrid residual
  feature learning with compact back-projection network. In: Proceedings of the
  IEEE International Conference on Computer Vision Workshops (2019)

\end{thebibliography}

\end{document}